\documentclass{PoS}

\graphicspath{{./figs/}}
\usepackage{graphicx}
\usepackage[hyphens]{url}
\usepackage{cite}
\usepackage{array}
\usepackage{amsmath}
\usepackage{multirow}
\usepackage{caption}
\usepackage{subcaption}
\usepackage{mdframed}
\usepackage{color}
\usepackage[usenames,dvipsnames]{xcolor}
\usepackage{listings}

\title{Code Optimization on Kepler GPUs and Xeon Phi}

\ShortTitle{Code Optimization on Kepler GPUs and Xeon Phi}

\author{Yong-Chull Jang, Hwancheol Jeong, Jangho Kim, Weonjong Lee, and
  \speaker{Jeonghwan Pak} \\ Lattice Gauge Theory Research Center, CTP, and
  FPRD, \\ Department of Physics and Astronomy, \\ Seoul National
  University, Seoul, 151-747, South Korea \\ E-mail: \email{wlee@snu.ac.kr}
}

\author{Yuree Chung \\ Hankuk academy of foreign studies, Yongin-si,
  Gyeonggi-do, 449-854, South Korea }

\abstract{
Kepler GTX Titan Black and Kepler Tesla K40 are still the best GPUs
for high performance computing, although Maxwell GPUs such as GTX 980
are available in the market.
Hence, we measure the performance of our lattice QCD codes using the
Kepler GPUs.
We also upgrade our code to use the latest CPS (Columbia Physics
System) library along with the most recent QUDA (QCD CUDA) library for
lattice QCD.
These new libraries improve the performance of our conjugate gradient
(CG) inverter so that it runs twice faster than before.
We also investigate the performance of Xeon Phi 7120P coprocessor.
It has similar computing power with the Kepler GPUs in principle.
However, its performance for our CG code is significantly inferior to
that of the GTX Titan Black GPUs at present.
}

\FullConference{The 32nd International Symposium on Lattice Field Theory,\\
		23-28 June, 2014\\
		Columbia University New York, NY}

\begin{document}


\section{Introduction}

We reported the performance of NVIDIA Kepler GTX Titan in Lattice 2013
\cite{Jeong:2013PoS}.
Since Lattice 2013, NVIDIA has released new Kepler GPUs such as GTX
Titan Black and Tesla K40, which show even better computing
performance than Maxwell GPUs.
We measure the performance of GTX Titan Black and Tesla K40m.
We also upgrade our code by adopting QUDA (library for QCD on GPU) so
that we can get a better performance from the Kepler GPUs.

We reported the performance of single Intel Xeon Phi 5110P card in
Lattice 2013 \cite{Jeong:2013PoS}.
In 2014, we perform the same test on a multiple Xeon Phi 7120P system.
Here we report its performance. 
We also find a problem on a parallel computing system with multiple
Xeon Phi 7120P cards connected via infiniband network.
We will describe this in detail in Section \ref{sec:phi}.


\section{GTX Titan Black}


\begin{table}[h]
  \begin{tabular}{|c|>{\centering\arraybackslash}m{1.3cm}
      |>{\centering\arraybackslash}m{1.3cm}
      |>{\centering\arraybackslash}m{1.3cm}
      |>{\centering\arraybackslash}m{1.3cm}
      |>{\centering\arraybackslash}m{1.3cm}
      |>{\centering\arraybackslash}m{1.3cm}|}
    \hline
    Architecture & Fermi & \multicolumn{4}{c|} {Kepler} & Maxwell \\
    \hline
    & GTX 580 & Tesla K20X & Tesla K40 & GTX 780Ti & GTX TITAN Black & GTX 980 \\
    \hline
    SP TFLOPS & 1.58 & 3.95 & 4.29 & \color{blue} 5.04 & \color{blue} 5.1 &
    4.61 \\
    \hline
    DP TFLOPS & 0.20 & \color{blue} 1.31 & \color{blue} 1.43 & 0.21 &
    \color{blue} 1.3 & 0.14 \\
    \hline
    Memory Size {\scriptsize (GB)} & 1.5 & \color{blue} 6 & \color{blue}
    12 & 3 & \color{blue} 6 & 4 \\
    \hline
    Memory Bandwidth {\scriptsize (GB/sec)} & 192.4 & 250 & 288 & \color{blue} 336 & \color{blue}
    336 & 224 \\
    \hline
    L2 cache {\scriptsize (MB)} & 7.7 & 1.5 & 1.5 & 1.5 & 1.5 & \color{blue} 2 \\
    \hline
    Price {\scriptsize (USD)} & N/A & 3,000 & 4,000 & 800 & 1,300 & 700 \\
    \hline
  \end{tabular}
  \caption{\label{tbl:spec}
    chip and memory specifications of recent NVIDIA GPUs
  }
\end{table}
In Table \ref{tbl:spec}, we present chip and memory specification of
high-end NVIDIA GPUs.
Although Maxwell GPUs are cheap, they have very poor double precision
performance and lower memory bandwidth than GTX Titan Black and Tesla
K40.
Hence, GTX Titan Black and Tesla K40 are still the best GPUs for
high performance computing at present.

\begin{figure}[tbhp]
  \centering
  \vspace*{-5mm}
  \includegraphics[width=0.6\linewidth]{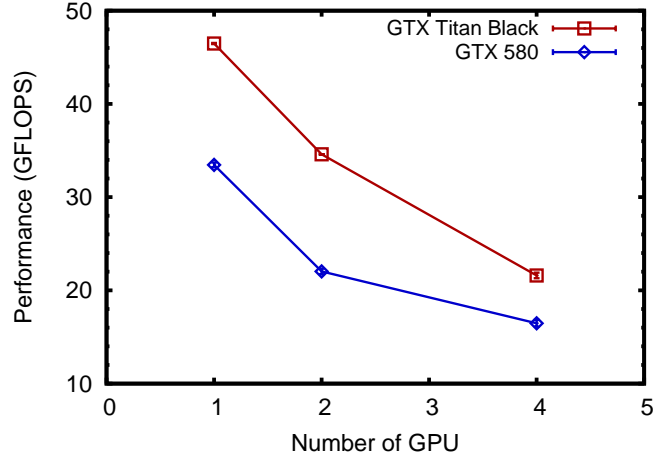}
        
  \caption{CG performance in the unit of giga flops. The data are
    obtained by running the CG code 10 times on the $20^3 \times 64$
    MILC asqtad lattice at $a \cong 0.12\;$fm. }
  \label{fig:cg_580_titanB_old}
\end{figure}
In Fig.~\ref{fig:cg_580_titanB_old}, we present the performance of our
conjugate gradient (CG) inverter by Fermi GTX 580 and Kepler GTX Titan
Black.
In the CG code, we use mixed precision algorithm, in which most of the
calculations are performed in single precision.
In this plot, two-GPU data points use the PCI bus, and four-GPU data
points use the PCI bus and infiniband network through MPI.
Hence, this plot shows that GTX Titan Black outperforms GTX 580 by
a factor of 1.3--1.6 regardless of the network environment.
Details of the plot are summarized in Table
\ref{tab:cg_580_titanB_old}.
\begin{table}
  \centering  
  \begin{tabular}{|>{\centering\arraybackslash}m{2.0cm}
      |>{\centering\arraybackslash}m{3.0cm}
      |>{\centering\arraybackslash}m{3.0cm}
      |>{\centering\arraybackslash}m{2.0cm}|}
    \hline
    \multirow{2}{*}{\# of GPU} & \multicolumn{2}{c|}{Performance (GFLOPS )}
    & \multirow{2}{*}{Ratio} \\
    \cline{2-3}
    & GTX 580 & GTX Titan Black & \\
    \hline
    1 & 33.46(26) & 46.48(6) & 1.39(1) \\
    \hline
    2 & 22.03(30) & 34.59(2) & 1.57(2) \\
    \hline
    4 & 16.47(36) & 21.60(30) & 1.31(1) \\
    \hline
  \end{tabular}
  \caption{ Results of CG performance in
    Fig.~\protect\ref{fig:cg_580_titanB_old}.  The ratio is taken
    between GTX Titan Black and GTX 580.
  } \label{tab:cg_580_titanB_old}
\end{table}

The performance includes data transfer between GPUs.
While GTX 580 supports PCI express 2.0, GTX Titan Black supports PCI
express 3.0 which provides twice larger bandwidth than the former.
We guess that the performance increase of GTX Titan Black with two
GPUs comes in part from the speed-up in the PCI bus.

\begin{figure}[tbhp]
  \centering
  \vspace*{-5mm}
  \includegraphics[width=0.6\linewidth]{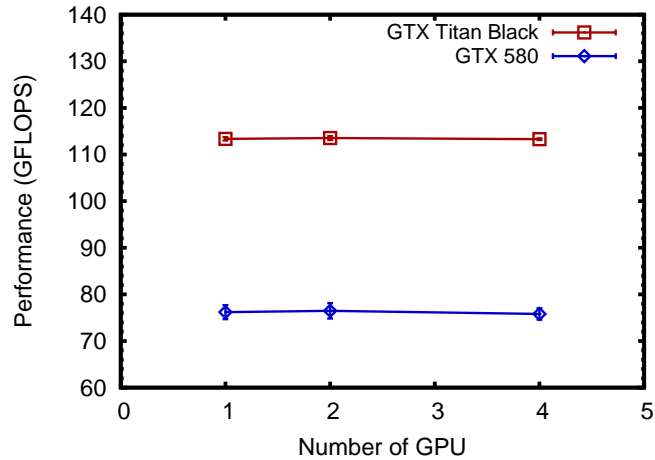}
  \caption{Performance of one-color four fermion operator part of the
    NPR production code in the unit of giga flops. The data are
    obtained by taking the average of 10 times running the code on the
    $20^3 \times 64$ MILC asqtad lattice at $a \cong 0.12\;$fm. }
  \label{fig:npr_580_titanB_old}
\end{figure}

Figure~\ref{fig:npr_580_titanB_old} shows the performance of the
one-color four fermion operator calculation of Eq.~\eqref{eq:npr_ff1c}
in our non-perturbative renormalization (NPR) production code on GTX
580 and GTX Titan Black.
\begin{align}  \label{eq:npr_ff1c}  
  O^{f_1 f_2 f_3 f_4}_{i;I} (z) =& \overline{\chi}^{f_1}_{i;c_1}(z_A)
  \overline{(\gamma_{S_1} \otimes \xi_{F_1})}_{AB} \chi^{f_2}_{i;c_2}(z_B)
  \times \overline{\chi}^{f_3}_{i;c_3}(z_C) \overline{(\gamma_{S_2} \otimes
    \xi_{F_2})}_{CD} \chi^{f_4}_{i;c_4}(z_D) \nonumber \\ & \times
         [U_{i;AD}]_{c_1 c_4} (z) [U_{i;CB}]_{c_3 c_2} (z)
\end{align}
where $\chi$ and $\overline{\chi}$ are staggered quark fields, and
$U_{i;AD} (z)$ is the gauge link averaged over all the shortest paths
between $A$ and $D$ at the hypercubic coordinate $z$.
Here, $S_1$ and $S_2$ represent the spins of the bilinears, and
$F_1$ and $F_2$ the tastes of the bilinears.
This part of the NPR code occupies about 98\% of the total running 
time.
Hence, this is the most dominant part in the NPR production code.
All of the calculation in the NPR code is done in double precision.
Therefore, we can measure the performance of the GPUs in double
precision floating point calculation using the NPR code.
As one can see in Fig.~\ref{fig:npr_580_titanB_old}, the double
precision performance is quite flat regardless of the network
environment because we exclude the network communication part (which
is negligibly small compared with the floating point calculation of
the GPUs) from this estimate.
\begin{table}[bhtp]
  \centering  
  \begin{tabular}{|>{\centering\arraybackslash}m{2.0cm}
      |>{\centering\arraybackslash}m{3.0cm}
      |>{\centering\arraybackslash}m{3.0cm}
      |>{\centering\arraybackslash}m{2.0cm}|}
    \hline
    \multirow{2}{*}{\# of GPU} & \multicolumn{2}{c|}{Performance (GFLOPS )}
    & \multirow{2}{*}{Ratio} \\
    \cline{2-3}
    & GTX 580 & GTX Titan Black & \\
    \hline
    1 & 76.19(150) & 113.36(35) & 1.49(2) \\
    \hline
    2 & 76.47(165) & 113.53(42) & 1.48(3) \\
    \hline
    4 & 75.80(122) & 113.29(19) & 1.49(2) \\
    \hline
  \end{tabular}
  \caption{ Results of NPR code performance in
    Fig.~\protect\ref{fig:npr_580_titanB_old}. The ratio is taken between
    GTX Titan Black and GTX 580.  } \label{tab:npr_580_titanB_old}
\end{table}

We expect that GTX Titan Black, in principle, outperforms GTX 580 by a
factor of $1.3/0.2 \approx 6.5$ in the double precision calculation.
However, we obtain only 1.5 times throughput rate in practice as one
can see in Table \ref{tab:npr_580_titanB_old}.
The main reason is that the memory bandwidth increase only by a factor
of 1.75 in the case of GTX Titan Black.
Therefore, the NPR code can not fetch the data fast enough to use the
full capability of the double precision calculation of GTX Titan
Black.
Our conundrum is that we expect to obtain still a gain of 1.75 (= the
maximum saturation limit of memory bandwidth) in the NPR code but have
gotten only the gain of 1.5 in practice.
This needs further investigation in near future.

To achieve the highest performance of the GTX Titan Black and other
Kepler GPUs, it is necessary to optimize the code as follows: (1) tune
thread and block scheduling, (2) increase CGMA (compute to global
memory access) ratio, (3) improve memory usage by new features such as
read-only data cache and warp shuffle, and (4) reduce unwanted
communications between GPUs and CPUs by applying direct parallelism
and GPU Direct \cite{Jeong:2013PoS}.
%




\section{CPS \& QUDA}
Recently we have upgraded the CPS (Columbia Physics System) library 
from version 4.2 to version 5.0.
The main advantage of the upgrade is that it allows us to use the QUDA
library which is a library for lattice QCD based on CUDA
\cite{web:quda}.
The QUDA library that we use for the performance test and production
jobs is version 0.5.1.
The QUDA library provides highly optimized version of the CG (conjugate
gradient) inverter and its variations such as BiCG (bi-conjugate
gradient) and PCG (preconditioned conjugate gradient) for the Nvidia
GPUs.
At present, it is designed to give the highest performance for the
Kepler family of Nvidia GPUs.
In addition, it also finds the optimized running parameters for the
system, dynamically and automatically, for any given system
environment.
It also supports the mixed precision algorithm for the CG inverter.

The latest CPS library (version 5.0) does not have an 
interface to call the CG inverter for staggered quarks in the
QUDA library.
Hence, recently we have added this interface to the CPS library
to run the staggered CG inverter of the QUDA library. 

\begin{figure}[t!]
  \centering
  \vspace*{-5mm}
  \includegraphics[width=0.6\linewidth]{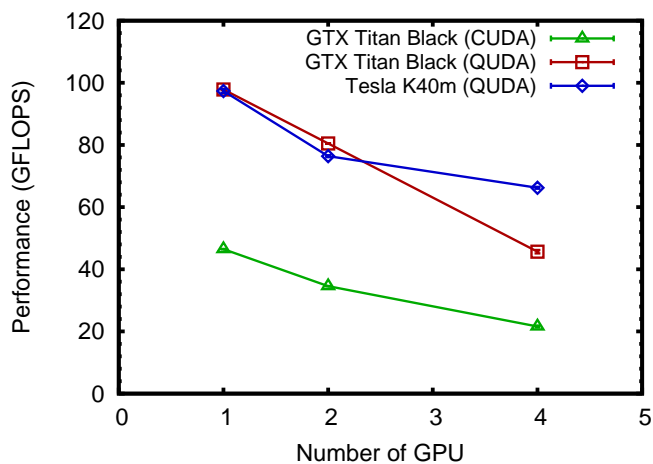}
        
  \caption{CG performance in the unit of giga flops. Here
    \texttt{CUDA} means the old version 4.2 of CPS library, and
    \texttt{QUDA} means the new version 5.0 of CPS library compiled
    with QUDA library. We use the same method as in
    Fig.~\protect\ref{fig:cg_580_titanB_old}. }
  \label{fig:cg_quda}
\end{figure}

In Fig.~\ref{fig:cg_quda}, we present the performance of the new CPS
library (version 5.0) compiled with the QUDA (version 0.5.1) library
(brown square symbols), and that (green triangle symbols) of the old
CPS library (version 4.2) compiled with the CUDA library (version
6.0).
In the case of the single GPU test, the new CPS library with QUDA
accelerate the speed of the CG inverter by a factor of 2.1.
In the case of two GPU test, the results of GTX Titan Black is obtained
using the PCI bus (bandwidth = 16 giga bytes per second), but that of
Tesla K40m is obtained using the infiniband FDR (fourteen data rate)
network (bandwidth = 56 giga bits per second).
In the case of four GPU test, the results of GTX Titan Black is
obtained using the infiniband QDR (quad data rate) network (bandwidth
= 32 giga bits per second), and that of Tesla K40m is obtained using
the infiniband FDR network.
In Table \ref{tab:cg_quda}, we summarize details of Fig.~\ref{fig:cg_quda}.
This shows that the gain in the performance of the CG inverter is
2.11 even in the infiniband network environment.  
\begin{table}[h!]
  \centering  
  \begin{tabular}{|>{\centering\arraybackslash}m{1.5cm}
      |>{\centering\arraybackslash}m{3.0cm}
      |>{\centering\arraybackslash}m{3.0cm}      
      |>{\centering\arraybackslash}m{2.0cm}
      |>{\centering\arraybackslash}m{3.0cm}|}
    \hline
    \multirow{2}{*}{\# of GPU} & \multicolumn{3}{c|}{Performance (GFLOPS )}
    & \multirow{2}{*}{Ratio} \\
    \cline{2-4}
    & Titan Black (old) & Titan Black (new) & K40m (new) & \\
    \hline
    1 & 46.48(6) & 97.86(21) & 97.37(60) & 1 : 2.11 : 2.09 \\
    \hline
    2 & 34.59(2) & 80.48(6) & 76.40(32) & 1 : 2.33 : 2.21 \\
    \hline
    4 & 21.60(30) & 45.66(55) & 66.24(30) & 1 : 2.11 : 3.07 \\
    \hline
  \end{tabular}
  \caption{ Results of CG performance in Fig.~\protect\ref{fig:cg_quda}.
  } \label{tab:cg_quda}
\end{table}
%
%
%




\section{Xeon Phi Coprocessor}
\label{sec:phi}
\begin{figure}[t!]
  \centering
  \vspace*{-5mm}
  \includegraphics[width=0.6\linewidth]{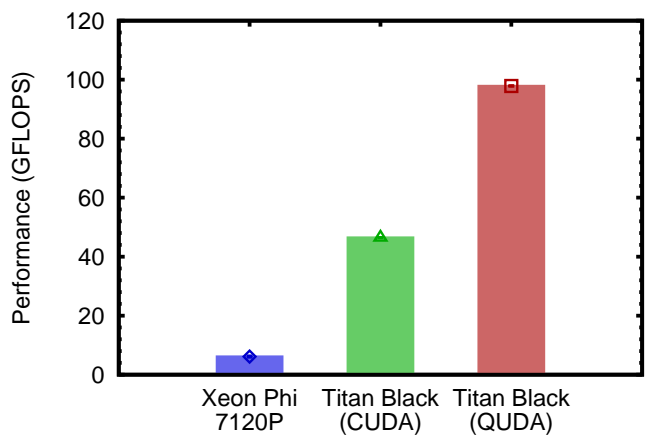}
        
  \caption{CG performance in the unit of giga flops. We use the same method
    as in Fig.~\protect\ref{fig:cg_580_titanB_old}.}
  \label{fig:cg_7120p_titanB}
\end{figure}

In Table \ref{tbl:spec_phi}, we present the specification of Xeon Phi
coprocessors and compare it with NVIDIA Kepler GPUs.
A Xeon Phi coprocessor has as much core performance and as large
memory bandwidth as the high-end NVIDIA Kepler GPUs. 
Hence, we expect that Xeon Phi can give, in principle, the same
performance as GTX Titan Black or Tesla K40.
\begin{table}[h]
  \centering
  \begin{tabular}{|c|>{\centering\arraybackslash}m{2.0cm}
      |>{\centering\arraybackslash}m{2.0cm}
      |>{\centering\arraybackslash}m{2.0cm}
      |>{\centering\arraybackslash}m{2.0cm}|}
    \hline & \multicolumn{2}{c|}{Intel Xeon Phi} &
    \multicolumn{2}{c|}{NVIDIA GPU} \\
    \hline
    & 5110P & 7120P & Tesla K40 & GTX Titan Black \\
    \hline
    \# of Cores & 60 & 61 & 2880 & 2880 \\
    \hline
    Core Clock (MHz) & 1053 & 1333 & 745 & 889 \\
    \hline
    SP TFLOPS & 2.02 & 2.44 & 4.29 & 5.1 \\
    \hline
    DP TFLOPS & \color{blue} 1.01 & \color{blue} 1.22 & \color{blue} 1.43 &
    \color{blue} 1.3 \\
    \hline
    Memory Size {\scriptsize (GB)} & \color{blue} 8 & \color{blue} 16 &
    \color{blue} 12 & 6 \\
    \hline
    Mem. Bandwidth {\scriptsize (GB/s)} & \color{blue} 320 & \color{blue}
    352 & 288 & 336 \\
    \hline
  \end{tabular}
  \caption{specification comparison between Intel Xeon Phi coprocessors and
    NVIDIA Kepler GPUs}
  \label{tbl:spec_phi}
\end{table}
The Xeon Phi coprocessor supports 512 bit SIMD (single instruction
multiple data) operations so that it can compute 16 single precision
calculations simultaneously.
Hence, the actual performance of Xeon Phi system highly depends on the
vectorization of the code.
Figure~\ref{fig:cg_7120p_titanB} shows the performance of the CG code on
Xeon Phi 7120P and GTX Titan Black.
Results are summarized in Table \ref{tab:cg_7120p_titanB}.
Here, note that we have not done the proper vectorization of our CG
code for Xeon Phi 7120P yet.
In this case, we report that the performance of our CG code for Xeon
Phi 7120P is much less than that for GTX Titan Black by a factor of
1/16.
\begin{table}[t!]
  \centering
  \vspace*{-5mm}
  \begin{tabular}{|>{\centering\arraybackslash}m{6.0cm}
      |>{\centering\arraybackslash}m{5.0cm}
      |>{\centering\arraybackslash}m{2.0cm}|}
    \hline
    Device & Performance (GFLOPS ) & Ratio \\
    \hline
    Xeon Phi 7120P & 6.14(30) & 1 \\
    \hline
    GTX Titan Black (CUDA) & 46.48(6) & 7.57(36) \\
    \hline
    GTX Titan Black (QUDA) & 97.86(21) & 15.94(74) \\
    \hline
  \end{tabular}
  \caption{ Results of CG performance in
    Fig.~\protect\ref{fig:cg_7120p_titanB}.
  } \label{tab:cg_7120p_titanB}
\end{table}

We find two major reasons for the poor performance of Xeon Phi.
First, our code is not fully vectorized to use 512 bit SIMD operations
through the automatic vectorization option of the Intel compiler.
To do this, we need to modify the code such that the code feeds the data
in the unit of 16 floating point numbers to the cores.
In order to achieve this properly, we may have to write the code in
the assembly language.
Second, we assign 200 MPI processes per Xeon Phi, and the performance
of each core decreases due to a big loss in the data distribution
speed of Xeon Phi.
Part of this problem comes from the poor software support on the MPI
library.
Hence, we plan to use OpenMP to distribute the data to the cores
because the OpenMP library supports the shared memory so that it can
alleviate the bottleneck problem in data distribution somewhat.
There has been a report on the success in implementing OpenMP and SIMD
vectorization to achieve as high performance as Tesla K20m in
Ref.~\cite{web:Joo}.


%
We also tested the scalability of the multiple Xeon Phi system.
If we use multiple Xeon Phi 7120Ps through MPI, the performance
of the system drops abruptly, and sometimes the system even crashes
completely.
We guess that there must be some bug of the native method on the
multiple Xeon Phi system with the infiniband network.
Hence, we plan to investigate the offload method.
%



\section{Conclusion}
GTX Titan Black improves the performance of our lattice QCD GPU codes
by 40\% $\sim$ 60\% compared with GTX 580 without any optimization
specific to Kepler GPUs.
On top of that, we achieved 2.1 times more performance by adopting the
latest CPS library with the QUDA library for staggered quarks.
Meanwhile, in order to obtain a comparable performance from Xeon Phi
system, we should optimize our code further using the vectorization
and OpenMP, which needs further investigation in near future.



\section{Acknowledgement}

The research of W.~Lee is supported by the Creative Research
Initiatives program (No.~2014001852) of the NRF grant funded by the
Korean government (MSIP).
This work was supported by SNU Undergraduate Research Program.
This work was supported by Seoul Science High School R\&E program.
W.~Lee would like to acknowledge support from the KISTI supercomputing
center through the strategic support program [No.~KSC-2013-G2-005]
with much gratitude.
%



\bibliographystyle{JHEP}
\bibliography{ref}


\end{document}